\documentclass{JHEP3}
\usepackage{amsmath}
\usepackage{axodraw}
\usepackage{cite}
\usepackage{epsfig}
\usepackage{xspace}
\usepackage{graphics}
\usepackage[utf8]{inputenc}
\usepackage{relsize}

\newcommand{\eqRef}[1]{eq.~\eqref{#1}\xspace}
\newcommand{\eqsRef}[1]{eqs.~\eqref{#1}\xspace}

\newcommand{\pythia}{P{\smaller YTHIA}\xspace}
\newcommand{\herwigpp}{H{\smaller ERWIG++}\xspace}
\newcommand{\herwig}{H{\smaller ERWIG}\xspace}
\newcommand{\dipsy}{{\smaller DIPSY}\xspace}

\def\done#1{}
\def\pmb#1{{\mbox{\boldmath$#1$}}}
\def\text{\mathrm}

\keywords{QCD, Jets, Parton Model, Phenomenological Models}
\preprint{arXiv:yymm.nnnn [hep-ph]\\
LU-TP 11-12\\
CERN-PH-TH-2011-059\\
MCnet-11-09\\
\today}

\skip\footins = 1\bigskipamount plus 2pt minus 4pt                              

\title{Correlations in double parton distributions at small
  $x$\footnote{Work supported in parts by the EU Marie Curie RTN MCnet
    (MRTN-CT-2006-035606), and the Swedish research council (contracts
    621-2008-4252 and 621-2009-4076).}}

\author{Christoffer Flensburg$^1$, Gösta Gustafson$^1$, Leif Lönnblad$^{1,2}$ and Andras Ster$^{1,3}$\\
  $^1$Dept.~of Astronomy and Theoretical Physics, Lund University, \\
  Sölvegatan 14A, Lund, Sweden\\
  $^2$CERN Theory Department, Geneva, Switzerland\\
   $^3$MTA KFKI-RMKI Department of Theoretical Physics, \\
   Konkoly Thege 29-33, H-1121 Budapest, Hungary.}
  
 \abstract{We present a dynamical study of the double parton distribution in 
impact parameter space, which enters into the double scattering cross section 
in hadronic collisions. This distribution is analogous to the generalized
parton densities in momentum space. We use the Lund Dipole Cascade model,
presented in earlier articles, which is based on BFKL evolution including
essential higher order corrections and saturation effects. As result we
find large correlation effects, which break the factorization of the
double scattering process. At small transverse separation we see the
development of "hot spots", which become stronger with increasing $Q^2$.
At smaller $x$-values the distribution widens, consistent with the
shrinking of the diffractive peak in elastic scattering. The dependence on
$Q^2$ is, however, significantly stronger than the dependence on $x$,
which has implications for extrapolations to LHC, \emph{e.g.} for results for 
underlying events associated with the production of new heavy particles.

}

\begin{document}
 
\sloppy
 
\section{Introduction}
\label{sec:intro}

Analyses of high energy $pp$ collisions show that multiple hard parton
collisions are quite common. Four-jet events, in which the jets
balance each other pairwise, were observed at $\sqrt{s}=63$~GeV by the
AFS collaboration at the CERN ISR \cite{Akesson:1986iv}, and at the
Tevatron events with four jets, or with three jets + $\gamma$, have been
observed by the CDF \cite{Abe:1993rv,Abe:1997xk} and D0 
\cite{Abazov:2002mr,Abazov:2009gc}
collaborations.  These results also show that the hard subcollisions
are correlated, and double interactions occur with a larger probability 
than expected for uncorrelated hard interactions. 

Knowing the cross section for double subcollisions is important for an
estimate of the underlying event. A good understanding of the 
correlations is therefore also essential for the interpretation of signals for 
new physics at the LHC. 
The aim of this paper is to study what kind of dynamical correlations 
in momentum space and impact parameter space
are to be expected at higher energies, as a result of parton evolution
to small $x$ and of saturation effects. 

Correlations which follow from energy and flavour conservation have been
discussed  \emph{e.g.} in 
refs.~\cite{Kuti:1971ph,Shelest:1982dg,Sjostrand:1987su}, and
from an impact parameter picture in ref.~\cite{Chou:1985sn}. Nontrivial
correlations which are consequences of DGLAP evolution have been studied in
refs.~\cite{Shelest:1982dg,Snigirev:2003cq,Gaunt:2009re}. In ref.~\cite{Snigirev:2010tk} it is shown that this implies an increasing
correlation for higher $Q^2$. However, in these
references the connection between correlations in momentum and $b$-space
have not been studied. Correlation effects also follow from fluctuations in the
number of particles in the parton cascades, which is discussed in 
refs.~\cite{Kuti:1971ph,Shelest:1982dg}. For related problems in connection
with production of two heavy gauge bosons, see \emph{e.g.} 
refs.~\cite{Diehl:2011tt, Gaunt:2011xd}.

In the model by Sj\"{o}strand and van Zijl \cite{Sjostrand:1987su},
later modified in ref.~\cite{Sjostrand:2004pf} and
implemented in the \pythia event generator
\cite{Sjostrand:2004ef,Sjostrand:2007gs,Sjostrand:2006za}, it is assumed that
the dependence of the double parton density on the kinematic variables
($x$ and $Q^2$) and the separation in impact parameter space
($\mathbf{b}$) factorizes. The correlation is described by a
distribution in the separation $\mathbf{b}$, where a denser central
region implies that central collisions have on average more hard
interactions, and peripheral collisions fewer. The relative shape of
this distribution is kept constant, but the width is scaled with
energy, to accommodate the growth of the total (non-diffractive) and
the multiple interaction cross sections with increasing energy. For 
the dependence on $x$ and $Q^2$ effects of energy and flavour conservation 
are taken into account. In a recent modification of the model, the shape is
also varying with $x$~\cite{Corke:2011yy}, in a way which gives smaller 
correlations for smaller $x$. At a fixed energy, higher $Q^2$ is
related to larger $x$, which implies that the production of heavy mass objects
is more common in high multiplicity events. Similar
approaches are also implemented in the \herwigpp event generator
\cite{Bahr:2008dy,Bahr:2009ek,Bahr:2008pv}.

While it is straight forward to compare the models in \pythia and
\herwig with experimental data on double parton scattering, it is not
trivial to disentangle exactly the sources of possible correlations in
the underlying models. This makes it difficult to extrapolate to
higher energy and/or higher $Q^2$. Here we will instead use a detailed
dynamical model for parton evolution, for a study of the correlations
between gluons inside hadrons.

In a series of papers we have presented a model based on BFKL
\cite{Kuraev:1977fs,Balitsky:1978ic} evolution and saturation, which
well reproduces data on total, elastic and diffractive cross sections
in DIS and $pp$ collisions
\cite{Avsar:2005iz,Avsar:2006jy,Avsar:2007xg,Flensburg:2008ag,Flensburg:2010kq}.
It is based on Mueller's dipole cascade
model~\cite{Mueller:1993rr,Mueller:1994jq,Mueller:1994gb}, but
includes also non-leading effects from energy conservation and running
coupling, as well as confinement effects and saturation within the
evolution.  The model is implemented in a Monte Carlo (MC) program,
which makes it easy to study in detail the types of correlations and
fluctuations in $x$ and $Q^2$ as well as in impact parameter space, 
which are consequences of the parton evolution.
 
The correlations come from two different sources.
First the distribution in transverse separation will vary with energy 
and virtuality, as the impact
parameter profile widens at higher energy, but at the same time the
development of ``hot spots'' makes the effective double scattering profile
more narrow. Secondly, BFKL dynamics implies 
large fluctuations in the cascade evolution. This implies that double
scattering is more probable in events with a higher than average number of
partons. This effect is neglected in most phenomenological analyses, which 
are based on average parton densities.

The outline of this paper is as follows. First we will go through
experimental aspects and the theoretical framework for double parton
scattering in section \ref{sec:dps}. In section \ref{sec:dipsy} we
will then briefly describe the dipole model in the \dipsy Monte Carlo,
followed by a description in section \ref{sec:application} of how we
will use this model to investigate correlations in the double parton
densities. The results are presented in section \ref{sec:results}
followed by conclusions and outlook in section \ref{sec:conclusions}.

\section{Double Parton Scattering and Double Parton Distributions}
\label{sec:dps}

\subsection{Experimental results}

A measure of the correlation is given by the quantity
$\sigma_{\mathrm{eff}}$ defined by the relation
\begin{equation}
\sigma^D_{(A,B)} = \frac{1}{(1+\delta_{AB})} \frac{\sigma^S_A
  \sigma^S_B}{\sigma_{\mathrm{eff}}}.
\label{eq:sigmaeff}
\end{equation}
Here $\sigma^D_{(A,B)}$ is the cross section for the two hard
processes $A$ and $B$, $\sigma^S_A$ and $\sigma^S_B$ the corresponding
single inclusive cross sections, and $(1+\delta_{AB})^{-1}$ is a symmetry 
factor equal to 1/2 if $A=B$. If the hard interactions were
uncorrelated, $\sigma_{\mathrm{eff}}$ would be equal to the total
non-diffractive cross section. The CDF and D0 measurements give instead
$\sigma_{\mathrm{eff}} \sim 15$~mb, which thus is significantly
smaller.

As mentioned in the introduction, experimental results on double
parton interactions have been published from the ISR and the
Tevatron. (The UA2 collaboration at the CERN S$p\bar{p}$S collider has
presented a lower limit for $\sigma_{\mathrm{eff}}$.) The ISR
experiment \cite{Akesson:1986iv} studied events with four jets with a
summed transverse energy above $29\,$GeV in $pp$ collisions at
$\sqrt{s}=63$~GeV.  The observed rate corresponds to
$\sigma_{\mathrm{eff}}=5$~mb, indicating a very strong correlation
between the partons. As the sum of $E_\perp$ for the four jets is about
30~GeV, the typical $x$-values for the partons is about 0.25. These
partons must be dominated by valence quarks, and therefore not
representative for the partons involved at higher energies and smaller
$x$.

The CDF collaboration studied 4-jet events with $\sum p_{\perp
  \mathrm{jet}}>140$~GeV at $\sqrt s = 1.8\,$TeV \cite{Abe:1993rv},
which implies $x$-values $\sim$ 0.04. The result obtained was
$\sigma_{\mathrm{eff}}=12.1^{+10.7}_{-5.4}\,$mb. CDF also studied
events with $\gamma$+3jets \cite{Abe:1997xk}.  This gave a more clear
signal than the 4-jet events, and the result is
$\sigma_{\mathrm{eff}}=14.5\pm 1.7^{+1.7}_{-2.3}\,$mb. The photons had
$E_{\perp \gamma}>16\,$GeV, and the jets $E_{\perp jet}>5\,$GeV, and
thus the $x$-values are smaller. No dependence on the Feynman scaling
variable for the two pairs was observed within the ranges
$0.01<x_F(\gamma+\mathrm{jet})<0.4$ and
$0.002<x_F(\mathrm{dijet})<0.2$. We note, however, that the 4-jet and
$\gamma$+3jet processes are not equivalent, as the production of a
photon are particularly sensitive to quark distributions.

While the D0 4-jet signal is rather weak \cite{Abazov:2002mr}, this 
collaboration also measures a
clear signal for $\gamma$+3jets \cite{Abazov:2009gc}. The transverse momenta,
and thus the parton $x$-values, are larger than those in the CDF analysis;
$p_{\perp \gamma}>60\,$GeV and $p_{\perp \mathrm{jet}2}>15\,$GeV. The result
obtained is $\sigma_{\mathrm{eff}}=16.4\pm0.3\pm2.3\,$mb. D0 also
studied the dependence on $p_\perp$ of the second jet. Here
$\sigma_{\mathrm{eff}}$ dropped by 24\% when $p_\perp$ increased from 17.5 to
27.5~GeV, but this variation was fully within the errors of the measurement.

\subsection{Formalism}
\label{sec:formalism}

Following ref.~\cite{Gaunt:2009re} we define the ``double parton
distribution'' $\Gamma_{ij}(x_1,x_2,b;Q_1^2,Q_2^2)$, which describes
the inclusive density distribution for finding a parton of type $i$
with energy fraction $x_1$ at scale $Q_1^2$, together with a parton of
type $j$ with energy fraction $x_2$ at scale $Q_2^2$, and with the two
partons separated by a transverse distance $b$. The distributions
$\Gamma_{ij}$ are via a Fourier transformation related to the
``two-parton generalized parton distributions'' in transverse momentum
space, $D(x_1,x_2,Q_1^2,Q_2^2,\vec{\Delta})$, studied by Blok \emph{et al.}
\cite{Blok:2010ge}.

Assuming factorisation of two hard subprocesses $A$ and $B$, the cross
section for double scattering is given by\footnote{This result is true 
if the parton--parton scattering is
local in impact parameter space, which ought to be the case for hard
collisions with large $Q^2$-values.}
\begin{eqnarray}
\sigma^D_{(A,B)} = \frac{1}{1+\delta_{AB}} \sum_{i,j,k,l} \int 
{\Gamma_{ij}(x_1,x_2,b;Q_1^2,Q_2^2) 
\hat{\sigma}^A_{ik}(x_1,x_1') \hat{\sigma}^B_{jl}(x_2,x_2')} \nonumber \\
\times \Gamma_{kl}(x_1',x_2',b;Q_1^2,Q_2^2) dx_1 dx_2 dx_1' dx_2' d^2b.
\label{eq:sigmaAB}
\end{eqnarray}
Here $\hat{\sigma}$ is the cross section for a parton-level
subprocess.

The approximation used in the event generators \pythia and 
\herwig means that, apart from effects of energy and flavour
conservation (which are quite important), $\Gamma$ factorizes in the form
\begin{equation}
\Gamma_{ij}(x_1,x_2,b;Q_1^2,Q_2^2) \sim D^{i}(x_1, Q_1^2)D^{j}(x_2, Q_2^2)F(b;s).
\label{eq:factorisation}
\end{equation}
Here $D^{i}$ denotes the single parton
distribution for parton type $i$, and $F(b;s)$ describes the
distribution in the transverse separation between the two partons, and is 
assumed to be 
independent of $x_i$ and $Q_i^2$, and to be the same for quarks and
gluons. (As mentioned in the introduction, in a recent
option in \pythia, $F$ depends also on $x$~\cite{Corke:2011yy}.) 
The width is adjusted by tuning the model to
reproduce the total cross section and multiple interactions, and thus
depends on the collision energy $\sqrt{s}$, as indicated in
\eqRef{eq:factorisation}. 

As mentioned in the introduction, the relation in eq.~\eqref{eq:factorisation} is
not consistent with DGLAP evolution. Assume that the distribution of two gluons satisfies the factorization relation in~\eqref{eq:factorisation} for some specific values of $x_1, Q_1^2, x_2,Q_2^2$.
Evolving to higher virtualities, gluon 1 and gluon 2 can form separate
cascades by emitting new gluons. A pair of gluons can then be formed,
either
with one from each of these cascades, or by two gluons from the same
cascade. The latter contribution will then necessarily break the
factorization relation \cite{Gaunt:2009re}. Gaunt and Stirling
assume instead the weaker relation
\begin{equation}
\Gamma_{ij}(x_1,x_2,b;Q_1^2,Q_2^2) = D^{ij}(x_1,x_2;Q_1^2,Q_2^2) F^i_{j}(b).
\label{eq:factgaunt}
\end{equation}
The relations in eq.~\eqref{eq:sigmaAB} and \eqRef{eq:factorisation} or 
\eqref{eq:factgaunt} imply 
that the effective cross section in \eqref{eq:sigmaeff} is determined by the relation
\begin{equation}
\sigma_{\mathrm{eff}}  = \left[\int d^2b (F(b))^2 \right]^{-1}.
\label{eq:sigmaeff2}
\end{equation}

In the following we want to use the Lund Dipole Cascade model to study
the correlations and fluctuations, which follow from the parton
evolution in a proton, and see how well the approximations in
\eqsRef{eq:factorisation} and \eqref{eq:factgaunt} are satisfied.  We
note, however, that as our model is based on the BFKL evolution, it
contains only gluons and should only be trusted at small $x$-values
where the gluons dominate.

We define the distribution $F(b;x_1,x_2,Q_1^2,Q_2^2)$ by the relation 
\begin{equation}
\Gamma(x_1,x_2,b;Q_1^2,Q_2^2) = D(x_1, Q_1^2)D(x_2, Q_2^2)
F(b;x_1,x_2,Q_1^2,Q_2^2).
\label{eq:defF}
\end{equation}
Thus $F$ is a density in transverse coordinate space $\mathbf{b}$,
which may depend on all four variables $x_1$, $x_2$, $Q_1^2$, and
$Q_2^2$, and which contains all information about correlations between
the two partons. In case \emph{e.g.} some kind of ``hot spots''
develop for small $x$ and/or large $Q^2$, this will show up as an
increase in $F$ for small $b$-values. With this
definition $F$ is related to the double scattering cross section via
\eqRef{eq:sigmaeff} and the relation
\begin{equation}
\sigma_{\mathrm{eff}}  = \left[\int d^2b F(b;x_1,x_2,Q_1^2,Q_2^2)
F(b;x_1',x_2',Q_1^2,Q_2^2) \right]^{-1},
\label{eq:sigmaefflund}
\end{equation}
with the constraint
\begin{equation}
s=\frac{Q_1^2}{x_1 x_1'}=\frac{Q_2^2}{x_2 x_2'}.
\end{equation}

For hard subcollisions at midrapidity we have $x_1\approx x_1'$ and
$x_2\approx x_2'$, and thus recover \eqRef{eq:sigmaeff2}, with the
difference that $\sigma_{\mathrm{eff}}$ may now depend on the
variables $x_i$ and $Q_i^2$ (with $Q_1^2/x_1^2=Q_2^2/x_2^2=s$).
We note, however, that for subcollisions away from midrapidity
we get different $x$-values in the
two $F$-distributions in \eqRef{eq:sigmaefflund}. This feature will be
further discussed in sec.~\ref{sec:results}.

\subsection{Correlations}

There are two different sources for correlations between the partons,
which both give contributions to $\sigma_{\mathrm{eff}}$.

\subsubsection{Distribution in impact parameter space}
\label{sec:pythiaimpact}

More central collisions are expected to have more hard subcollisions than
peripheral collisions. This feature causes correlations, which depend on the 
matter distribution within the colliding protons. To understand the effect of 
the matter distribution, we here study a simplified  version of the model 
implemented in the \pythia event generator. A discussion of the results in our
model, and a comparison with more refined versions of \pythia, are presented 
in sec.~\ref{sec:results}. 

Assume that the parton density inside a proton is
given by a Gaussian distribution $\rho\propto \exp(-r^2/a^2)$. Integrating
over the longitudinal coordinate $z$ gives a factor $\sqrt{\pi}\, a$
independent of the impact parameter $b=\sqrt{x^2+y^2}$. Thus the density in
impact parameter space is also given by 
\begin{equation}
\rho\propto \exp(-b^2/a^2).
\end{equation}
The two parton density is then given by 
\begin{equation}
F(b)\propto  \int d^2r \rho(\mathbf{r}) \rho(\mathbf{r}-\mathbf{b})
\propto  \exp(-b^2/2a^2).
\label{eq:rhoint}
\end{equation}
The $F$-distribution is normalized to 1, and has therefore a normalization
constant equal to $(\pi 2 a^2)^{-1}$. This implies that
\begin{equation}
\sigma_{\mathrm{eff}}=\left(\int d^2b\, F^2\right)^{-1}= 4\pi a^2.
\end{equation}

In \pythia it is assumed that for two colliding protons,
the average number of hard subcollisions is
proportional to the overlap of the two distributions. For a
collision at impact parameter $b$, this overlap is given by
\begin{equation}
\mathcal{O}(b)= \int d^2r \rho(\mathbf{r}) \rho(\mathbf{r}-\mathbf{b})
\propto  \exp(-b^2/2a^2).
\label{eq:overlap}
\end{equation}
We note that the integral in \eqRef{eq:overlap} is exactly the same
as the one determining $F$ in
\eqRef{eq:rhoint}. The distribution $F$ is normalized to 1, and we
therefore find 
\begin{equation}
F(b)=\mathcal{O}(b)/\int d^2b\, \mathcal{O}(b).
\end{equation}

Since in \pythia the average number of hard subcollisions at a fixed impact
parameter, $\bar{n}(b)$, is proportional to the overlap, $\mathcal{O}(b)$, it can
be written as
\begin{equation}
\bar{n}(b)\propto \mathcal{O}(b)\cdot \hat{\sigma},
\end{equation}
where the cross section for parton--parton collisions is approximated by 
$\hat{\sigma}\,\delta^{(2)}(\mathbf{b})$.
The number of subcollisions is assumed to be given by a Poisson distribution
with average $\bar{n}(b)$, and a non-diffractive event is obtained if
$n\neq 0$. This implies that the total non-diffractive cross section
is given by
\begin{equation}
\sigma_{\mathrm{ND}}=\int d^2b\, (1-e^{-\bar{n}(b)}),
\label{eq:sigmaND}
\end{equation}
and the average number of subcollisions is
\begin{equation}
\langle n\rangle=\frac{\int d^2b \,\bar{n}(b)}{\sigma_{\mathrm{ND}}}.
\label{eq:bar-n}
\end{equation}

In \pythia $\hat{\sigma}$ depends on a parameter $p_{\perp 0}$, which acts as
a smooth cutoff for small $p_\perp$ in
parton--parton scattering. We have therefore two adjustable parameters, 
$a$ and $p_{\perp 0}$, which can be used to fit $\sigma_{\mathrm{ND}}$ and
$\langle n\rangle$. A result from such a fit is presented in 
ref.~\cite{Corke:2011yy}, with the result $a\approx 0.48$~fm, giving
$\sigma_{\mathrm{eff}}\approx 28$~mb, at the Tevatron and 
$\sigma_{\mathrm{eff}}\approx 34$~mb at LHC. 

These values could be regarded as a kind of benchmarks. If the
evolution gives smaller regions with higher parton density, the width of the
$F$-distribution will be wider, and $\sigma_{\mathrm{eff}}$ will be smaller.
We here note that, although the single Gaussian fit
gives the proper number of subcollisions, the correlations appear to be too
week. More successful tunes to data have a distribution $\rho$ (and thus also
$\mathcal{O}$), which is given by a sum of two Gaussian. This enhances small
and large $b$-values, and gives a somewhat smaller $\sigma_{\mathrm{eff}}$.
In a recent modification to the model \cite{Corke:2011yy}, the width of the 
density $\rho$ is assumed to vary with $x$, which also has the effect of
enhancing small and large $b$-values increasing the correlations and reducing 
$\sigma_{\mathrm{eff}}$.

\subsubsection{Fluctuations in the parton cascade} 

Another source for correlations is coming from fluctuations within the
cascades, and affects $\sigma_{\mathrm{eff}}$ even if the double parton
distribution factorizes as in \eqRef{eq:factorisation}. 
This effect is discussed in refs.~\cite{Kuti:1971ph} and \cite{Shelest:1982dg},
but is normally neglected in phenomenological
analyses.  Double hard scattering is more likely in collisions with
protons that have more than the average number of partons.  A measure
of this effect is given by the integral of $F$ as follows:

The cascade evolution is a random process, which can lead to different
partonic states. We label the states by the parameter $n$, and the
parton distribution in state $n$ is denoted $D_n(x,Q^2)$. The
probability to obtain this state is denoted $P_n$, with $\sum
P_n=1$. For a fixed state $n$ the density of partons in a small interval $\delta
x_i$ around $x_i$ at resolution $Q_i^2$, is given by $D_n(x_i,Q_i^2)\delta
x_i$. As we have a single state, the existence of a parton at $x_1$ is
fixed, and it is not possible to define a correlation with the existence
of another parton at $x_2$. Therefore
$D_n(x_1,Q_1^2,x_2,Q_2^2)=D_n(x_1,Q_1^2)\cdot D_n(x_2,Q_2^2)$. Averaging
over all states $n$ we thus get
\begin{equation}
\int d^2b \,F(b)=\frac{\sum_n P_n\, D_n(x_1,Q_1^2)\, D_n(x_2,Q_2^2)}
   {\sum_n P_n D_n(x_1,Q_1^2) \cdot \sum_n P_n D_n(x_2,Q_2^2) }=
   \frac{\langle D(x_1,Q_1^2) D(x_2,Q_2^2)\rangle}
   {\langle D(x_1,Q_1^2)\rangle \langle D(x_2,Q_2^2)\rangle }.
\end{equation}
Here $\langle\ldots\rangle$ denotes an average over the different
cascades $n$. For the special case of $Q^2_1=Q^2_2$ and $x_1=x_2$, we
get
\begin{equation}
\int\, d^2b\, F(b)=\frac{\langle D^2(x,Q^2)\rangle}{\langle D(x,Q^2)\rangle^2}.
\end{equation}
Thus the integral of $F$
equals $1+V/\langle D(x,Q^2) \rangle^2$, with the variance $V$ equal
to the square of the width of the distribution.
We see that the fluctuations imply that the $F$-distribution is enhanced,
which also implies a larger correlation and a smaller $\sigma_{\mathrm{eff}}$.
When presenting our results in section~\ref{sec:results},
we will also include values for the integral of $F$, in order to
facilitate the interpretation of the results.  Note that in the
Gaunt-Stirling approximation in \eqRef{eq:factgaunt}, $F$ is
normalized to 1, and all correlation effects are included in the
function $D^{ij}(x_1,x_2;Q_1^2,Q_2^2)$.

\section{The Lund Dipole Cascade Model}
\label{sec:dipsy}

\subsection{Mueller's dipole cascade}

Mueller's dipole cascade model
\cite{Mueller:1993rr,Mueller:1994jq,Mueller:1994gb} is a formulation
of the leading logarithmic (LL) BFKL evolution in transverse
coordinate space.  Gluon radiation from the colour charge in a parent
quark or gluon is screened by the accompanying anti-charge in the
colour dipole. This suppresses emissions at large transverse
separation, which corresponds to the suppression of small $k_\perp$ in
BFKL.  For a dipole $(\pmb{x},\pmb{y})$ the probability per unit
rapidity ($Y$) for emission of a gluon at transverse position
$\pmb{z}$ is given by
\begin{eqnarray}
\frac{d\mathcal{P}}{dY}=\frac{\bar{\alpha}}{2\pi}d^2\pmb{z}
\frac{(\pmb{x}-\pmb{y})^2}{(\pmb{x}-\pmb{z})^2 (\pmb{z}-\pmb{y})^2},
\,\,\,\,\,\,\, \mathrm{with}\,\,\, \bar{\alpha} = \frac{3\alpha_s}{\pi}.
\label{eq:dipkernel1}
\end{eqnarray}
This emission implies that the dipole is split into two dipoles, which
(in the large $N_c$ limit) emit new gluons independently. The result is a
cascade, where the number of dipoles grows exponentially with $Y$.

In a high energy collision, the dipole cascades in the projectile and the 
target are evolved from their rest frames to the rapidities they will have
in the specific Lorentz frame chosen for the analysis.
The scattering probability between two
elementary colour dipoles with coordinates $(\pmb{x}_i,\pmb{y}_i)$ and
$(\pmb{x}_j,\pmb{y}_j)$ in the projectile and the target respectively, is 
given by $2f_{ij}$, where in the Born approximation
\begin{equation}
  f_{ij} = f(\pmb{x}_i,\pmb{y}_i|\pmb{x}_j,\pmb{y}_j) =
  \frac{\alpha_s^2}{8}\biggl[\log\biggl(\frac{(\pmb{x}_i-\pmb{y}_j)^2
    (\pmb{y}_i-\pmb{x}_j)^2}
  {(\pmb{x}_i-\pmb{x}_j)^2(\pmb{y}_i-\pmb{y}_j)^2}\biggr)\biggr]^2.
\label{eq:dipamp}
\end{equation}
The optical theorem then implies that the elastic amplitude for dipole $i$ 
scattering off dipole $j$ is 
given by $f_{ij}$. Summing over $i$ and $j$ gives the one-pomeron elastic 
amplitude
\begin{equation}
F=\sum f_{ij}.
\end{equation}

The growth in the number of dipoles also implies a strong growth for the 
interaction probability, but the total scattering probability is kept 
below 1 by the possibility
to have multiple dipole--dipole subcollisions in a single event.
In the eikonal approximation
the unitarized elastic amplitude is given by the exponentiated expression
\begin{equation}
T(\pmb{b})=1-e^{-F},
\label{tf-relationmueller}
\end{equation}  
and the total, elastic, and diffractive cross sections are given by
\begin{eqnarray}
d\sigma_{\text{tot}} / d^2b& =&\,\langle 2T \rangle, \nonumber\\
d\sigma_{\text{el}}/d^2b \,\,& =& \, \langle T\rangle^2, \nonumber\\
d\sigma_{\text{diff\, ex}}/d^2 b  & =&
\langle T^2 \rangle - \langle T \rangle ^2.
\label{eq:eikonalcross}
\end{eqnarray}

\subsection{The Lund dipole cascade model}

In refs.~\cite{Avsar:2005iz, Avsar:2006jy, Flensburg:2008ag} we describe a 
modification of Mueller's cascade model with the following features:
\begin{itemize}\itemsep 0mm
\item It includes essential NLL BFKL effects.
\item It includes non-linear effects in the evolution.
\item It includes effects of confinement.
\end{itemize}

The model also includes a simple model for the proton wavefunction,
and is implemented in a Monte Carlo simulation program called \dipsy.
As discussed in the cited references, the model is able to describe a
wide range of observables in DIS and $pp$ scattering, with very few
parameters.

\subsubsection{NLL effects}

The NLL corrections to BFKL evolution have three major sources
\cite{Salam:1999cn}:

\vspace*{2mm}\emph {The running coupling:}\\
This is relatively easily included in a MC simulation process. The scale in
the running coupling is chosen as the largest transverse momentum in the
vertex \cite{Balitsky:2008zzb}.

\vspace*{2mm}\emph {Non-singular terms in the splitting function:}\\
These terms suppress large $z$-values in the individual parton
branchings, and prevent the daughter from being faster than her
recoiling parent.  Most of this effect is taken care of by including
energy-momentum conservation in the evolution. This is effectively
taken into account by associating a dipole of transverse size $r$ with
a transverse momentum $k_\perp = 1/r$, and demanding conservation of
the light-cone momentum $p_+$ in every step in the evolution. This
gives an effective cutoff for small dipoles, which also eliminates the
numerical problems encountered in the MC implementation by Mueller and
Salam \cite{Mueller:1996te}.  \vspace{2mm}

\vspace*{2mm}\emph{Projectile-target symmetry:}\\
This is also called energy scale terms, and is essentially equivalent
to the so called consistency constraint \cite{Kwiecinski:1996td}. This
effect is taken into account by conservation of both positive and
negative light-cone momentum components, $p_+$ and $p_-$.  The
treatment of these effects includes also effects beyond NLL, in a way
similar to the treatment by Salam in
ref.~\cite{Salam:1999cn}. Therefore the power
$\lambda_{\mathrm{eff}}$, determining the growth for small $x$, does
not turn negative for large values of $\alpha_s$.

\subsubsection{Non-linear effects and saturation}

As mentioned above, dipole loops (or equivalently pomeron loops) are
not included in Mueller's cascade model, if they occur within the
evolution.  They are only included if they are cut in the Lorentz
frame used in the calculations, as a result of multiple scattering in
this frame. The result is therefore not frame independent.  (The
situation is similar in the Color Glass Condensate
\cite{Iancu:2002xk,Iancu:2003xm,Gelis:2010nm} or the JIMWLK
\cite{Iancu:2000hn,JalilianMarian:1997jx} equations.) As for dipole
scattering the probability for such loops is given by $\alpha_s$, and
therefore formally colour suppressed compared to dipole splitting,
which is proportional to $\bar{\alpha}=N_c \alpha_s/\pi$. These loops
are therefore related to the probability that two dipoles have the
same colour. Two dipoles with the same colour form a quadrupole
field. Such a field may be better approximated by two dipoles formed
by the closest colour--anticolour charges. This corresponds to a
recoupling of the colour dipole chains, favouring the formation of 
small dipoles. We call this process a dipole
``swing''. The swing gives rise to loops within the cascades, and
makes the cross section frame independent to a good approximation.  We
note that a similar effect would also be obtained from gluon exchange
between the two dipoles.

In the MC implementation each dipole is assigned one of $N_C^2$ colour
indices, and dipoles with the same colour index are allowed to
recouple \cite{Avsar:2006jy}. The weight for the recoupling is assumed
to be proportional to $(r_1^2 r_2^2)/(r_3^2 r_4^2)$, where $r_1$ and
$r_2$ are the sizes of the original dipoles and $r_3$ and $r_4$ are
the sizes of the recoupled dipoles. Dipoles with the same colour are allowed
to  swing back and forth, which results in an equilibrium, where the smaller
dipoles have a larger weight. We note that in this formulation
the number of dipoles is not reduced, and the saturation effect is obtained
because the smaller dipoles have smaller cross sections. Thus in an
evolution in momentum space the swing would not correspond to an
absorption of gluons below the saturation line $k_\perp^2 = Q_s^2(x)$;
it would rather correspond to lifting the gluons to higher $k_\perp$
above this line.  Although this mechanism does not give an explicitly
frame independent result, MC simulations show that it is a very good
approximation.

\subsubsection{Confinement effects}

Confinement effects are included via an effective gluon mass, which
gives an exponential suppression for very large dipoles \cite{Avsar:2007xg}. 
This prevents the proton from growing too fast in transverse size, and is also
essential to satisfy Froisart's bound at high energies \cite{Avsar:2008dn}.

\subsubsection{Initial dipole configurations}

In DIS an initial photon is split into a $q\bar{q}$ pair, and
for larger $Q^2$ the wavefunction for a virtual photon can be determined 
perturbatively. The internal structure of the proton is, however,
governed by soft QCD, and is not 
possible to calculate perturbatively. In our model it is represented by
an equilateral triangle formed by three dipoles, and with a radius of
$3$~GeV$^{-1} \approx 0.6$~fm. The model should be used at low $x$,
and when the system is evolved over a large 
rapidity range the observable results depend only weakly on the exact 
configuration of the initial dipoles, or whether the charges are treated as 
(anti)quarks or gluons.

\section{Application to double parton distributions}
\label{sec:application}

In principle we could estimate the gluon density in the proton from the cross
section for $\gamma^*p$ collisions.  The photon would then be treated as a 
superposition of dipoles of varying sizes, with weights determined by QED.
However, in order to more easily isolate the correlations and
fluctuations in the proton, without the complications from the
additional fluctuations in the photon wave functions, we prefer instead to 
calculate the cross section for a dipole with a fixed transverse size.
The photon coupling to a $q\bar{q}$ pair contains a factor
\mbox{$(z(1-z))^{2-i}K_i^2(\sqrt{z(1-z)}\,Qr)$}. Here $K_i$ are generalized Bessel
functions, $r$ is the transverse separation
between the quark and the antiquark, $z$ is the fractional energy taken by the
quark, and the index $i$ is 1 for transverse and 0 for longitudinal
photons. Important contributions are obtained for $z\sim 0.5$,
and since the Bessel functions fall off exponentially when the argument is
larger than 1, characteristic $r$-values are given by the relation 
$r\approx 2/Q$.

To determine the parton distributions $D(x,Q^2)$ we thus calculate the
cross section for scattering of a single dipole of size $r= 2/Q$ colliding 
with a proton. 
The projectile dipole has a large cross section when colliding with equally 
large or larger dipoles in the proton, while the interaction with smaller
dipoles is suppressed. The dipoles in the proton are also 
connected to gluons with $k_\perp$ of the order $1/r$, and therefore the
$D$-distributions obtained correspond to the \emph{integrated} gluon
distributions. 

The double parton density $\Gamma(x_1,x_2,b;Q_1^2,Q_2^2)$ 
is in the same way proportional to the cross section, $\sigma^{(1,2)}$, for 
simultaneous scattering of two dipoles with size $r_i=2/Q_i$, separated by a 
distance $b$, and
colliding with a proton. Thus also $\Gamma$ is defined as an integrated 
density. From these relations we conclude that
the $F$-distribution is directly related 
to the ratio between the corresponding cross sections, and in an obvious
notation we have
\begin{equation}
F(b;x_1,x_2,Q_1^2,Q_2^2)\equiv\frac{\Gamma^{(1,2)}}{D^{(1)} D^{(2)}}
=\frac{\sigma^{(1,2)}}{\sigma^{(1)}\sigma^{(2)}}.
\end{equation}

In principle the double scattering cross section, and thus also $F$, depends 
in the dipole model on the three vectors, 
the dipole sizes $\mathbf{r}_1$, $\mathbf{r}_2$, and the distance between
them $\mathbf{b}$. This implies that the estimate of $\sigma_{\mathrm{eff}}$
should be given by a 6-dimensional integral over these variables.
For the results presented in the next section we have 
calculated $F$ keeping the separation between two gluons constant equal to 
$\mathbf{b}$, but averaging over the directions of the dipoles, $\mathbf{r}_1$
and $\mathbf{r}_2$. To check this approximation we have also calculated the 
result obtained when keeping the
distance between the centers of the dipole fixed. The results presented in
sec.~\ref{sec:results} show that the correlations obtained are insensitive to 
how the averages are taken.
 
\section{Results}
\label{sec:results}

In this section we show some results relevant for $pp$ collisions at
$\sqrt{s} \approx 1.5$ and 15~TeV, qualitatively representing Tevatron and LHC
energies. We calculate the cross sections for one or two dipoles
against a proton, where the rapidity separation between the projectile
and the target is given by $y=\ln(1/x_i) +\ln(Q_i/2)$. Thus for
subcollisions at midrapidity, with $y=\ln{\sqrt{s}}$, we have
$x=Q/(2\sqrt{s})$.

\subsection{Subcollisions at midrapidity}

\begin{figure}
\centering
\epsfig{file=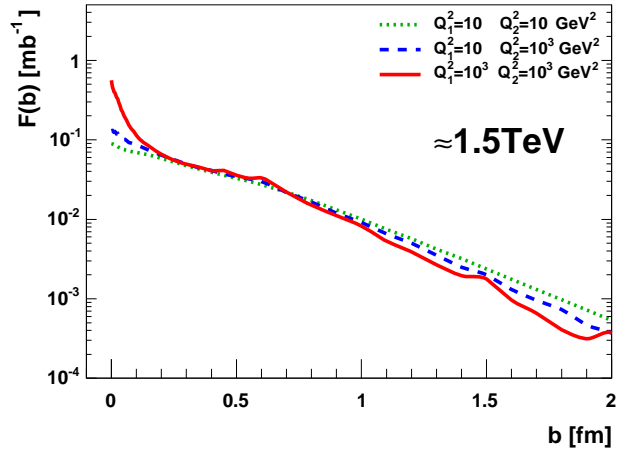,width=0.6\linewidth}
\caption{Correlation function $F(b)$ corresponding to central subcollisions at
  $\sqrt{s} \approx$ 1.5~TeV, and three combinations of $Q_1^2$ and $Q_2^2$. 
  The $x$-values corresponding to 
  $Q^2=10$ and 1000~GeV$^2$ are $10^{-3}$ and $10^{-2}$ respectively.  }
\label{fig:qdep1.5}
\end{figure}

\begin{figure}
\centering
\epsfig{file=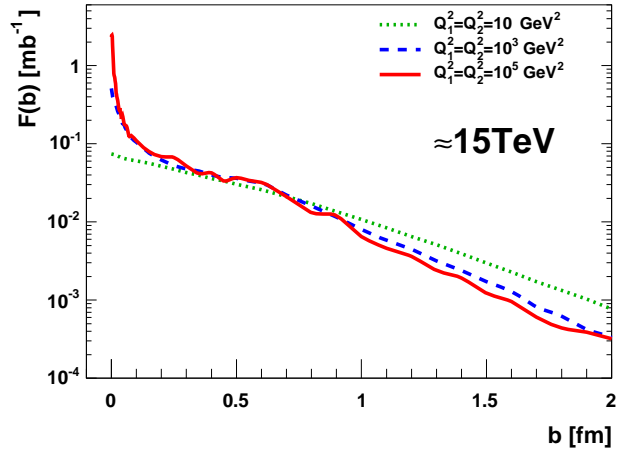,width=0.6\linewidth}
\caption{Correlation function $F(b)$ corresponding to central subcollisions at
  $\sqrt{s} \approx$ 15~TeV. The curves correspond to
  $Q_1^2=Q_2^2=10,\,10^3,\,\mathrm{and}\, 10^5\, \mathrm{GeV}^2$. 
  The corresponding $x$-values are $10^{-4}$, $10^{-3}$, and $10^{-2}$.}
\label{fig:qdep15}
\end{figure}

In figs.~\ref{fig:qdep1.5} and \ref{fig:qdep15} we show the
$Q^2$-dependence for $x$-values corresponding to central production at
a fixed energy.  Fig.~\ref{fig:qdep1.5} shows the result for
$\sqrt{s} \approx 1.5$~TeV and three combinations\footnote{The chosen values
  of $Q^2$ of $10$, $10^3$ and $10^5$~GeV$^2$, correspond to dipole
  sizes of $\approx 0.13$, $0.013$ and $0.0013$~fm respectively.} of
$Q_1^2$ and $Q_2^2$.  The corresponding $x$-values are given by
$x=Q/(2\sqrt{s})$.  Fig.~\ref{fig:qdep15} shows similar results for
$\sqrt{s} \approx 15\,\,\mathrm{TeV}$.

We see that the distribution is not well described by a Gaussian. Instead
there is an almost exponential dependence in the range
$0.2<b<0.6\,\mathrm{fm}$. For larger $b$-values the distribution drops faster
, and this effect is stronger for higher $Q^2$. The position of the break is
related to the size of the initial proton configuration in the model.
For small $b$-values a spike is developing,
growing stronger with increasing $Q^2$. This is associated with a faster drop
for larger $b$ for high $Q^2$. Results with one softer and one
harder subcollision lie (as expected) in between the results for two soft or
two hard collisions, but closer to the result for two softer subcollisions
(not shown for $\sqrt{s} \approx 15$~TeV).

\begin{figure}
\centering
\epsfig{file=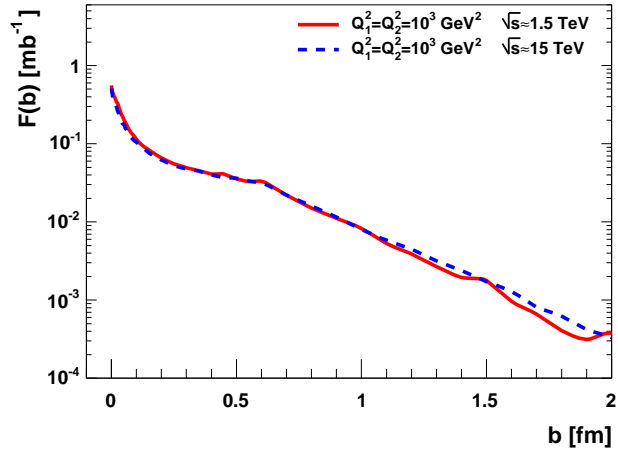,width=0.6\linewidth}
\caption{The energy dependence of the correlation functions $F(b)$. The
  curves show $F$ for $Q_1^2=Q_2^2=10^3$~GeV$^2$, solid line
   for $\sqrt{s} \approx 1.5$ and dashed line for $\sqrt{s} \approx 15$~TeV.}
\label{fig:sdep}
\end{figure}

In fig.~\ref{fig:sdep} we show the energy dependence by comparing the results 
for $Q^2= 10^3\,\mathrm{GeV}^2$ at the two different energies. We see here
that the peaks at small $b$ are almost identical, but at the higher energy
the distribution becomes a little wider, with a slightly larger tail out to 
large $b$-values.

The corresponding values for $\sigma_{\mathrm{eff}}=(\int d^2b\,F(b)^2)^{-1}$ 
and $\int d^2b\,F(b)$ were calculated numerically, and are
presented in table \ref{table:results}. The rather slow fall off for $F$ at
large $b$ gives a large $\sigma_{\mathrm{eff}}$ for small $Q^2$, but
the more narrow distributions for
higher $Q^2$ imply a very strong $Q^2$-dependence. At 15 TeV 
$\sigma_{\mathrm{eff}}$ drops by a factor 2 when $Q^2$ is changed from 10 to
10$^2$ GeV$^2$. For fixed $Q^2$ the wider distribution at higher energy, 
implies a slightly larger $\sigma_{\mathrm{eff}}$.
The variation with $\sqrt{s}$ is, however, much weaker than the variation
with $Q^2$. 

\begin{table}
\begin{center}
\begin{tabular}{|llll|r|r|}
\hline
\hline
\multicolumn{4}{|c|}{$Q_1^2,Q_2^2\mbox{~[GeV$^2$]},x_1,x_2$}    &$\sigma_{\mathrm{eff}}$~[mb]  &$\int F$\\
\hline
\multicolumn{4}{|c|}{ 1.5~TeV, midrapidity}  & &  \\
   10       &10      &0.001  &0.001         & 35.3  \hspace{3mm}      & 1.09  \\
   10      &$10^3$   &0.001  &0.01          & 31.0 \hspace{3mm}        & 1.07 \\
  $10^3$   &$10^3$   &0.01   &0.01          & 23.1  \hspace{3mm}      & 1.06  \\
\hline
\multicolumn{4}{|c|}{ 15~TeV, midrapidity}  & &   \\
   10       &10      &0.0001 &0.0001        & 40.4 \hspace{3mm}       & 1.11 \\
  $10^3$   &$10^3$   &0.001  &0.001         & 26.3  \hspace{3mm}  & 1.07  \\
  $10^3$   &$10^5$   &0.001  &0.01          & 24.2 \hspace{3mm}       & 1.05  \\
  $10^5$   &$10^5$   &0.01   &0.01          & 19.6 \hspace{3mm}       & 1.03   \\
\hline
\multicolumn{4}{|c|}{ 1.5~TeV, $y_{\mathrm{pair}\,2}=2.3$}  & &  \\
   10       &10      &0.001 &0.0001         &    &  \\
   10       &10      &0.001  &0.01         & 
   \raisebox{3mm}[0mm][0mm]{$\left.\vphantom{\frac{\int}{\int}}\right\}$~37.1} \hspace{3mm}   & 
   \raisebox{3mm}[0mm][0mm]{1.08} \\
\hline
\end{tabular}
\end{center}
\caption
{
Summary of results for $\sigma_{\mathrm{eff}}$ and corresponding
integrals of the double distribution functions. (The numerical uncertainties
are about 1\%.)
}
\label{table:results}
\end{table}

We also note that the effect of fluctuations in the
cascades, represented by the difference from 1 of $\int d^2b\,F$, is of the
order 5-10\% (contributing 10-20\% to $\sigma_{\mathrm{eff}}$), and is smaller
for large $Q^2$.

\subsection{Sub-collisions off midrapidity}
\label{sec:non-central}

\begin{figure}
\centering
\epsfig{file=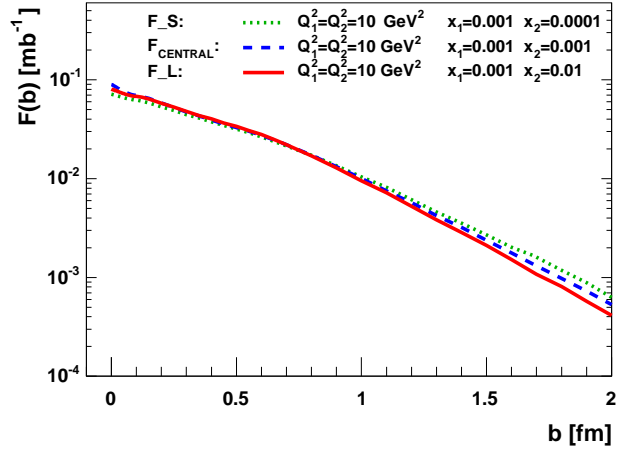,width=0.6\linewidth}
\caption{Correlation function $F(b)$ relevant for subcollisions at 
$y_{\mathrm{pair}\,2}=2.3$ for
  $\sqrt{s} \approx$ 1.5~TeV and $Q_1^2=Q_2^2=10$~GeV$^2$. 
$F_S \equiv F(b;\,0.001,\, 0.0001,\, 10,\, 10)$  and 
$F_L\equiv F(b;\,0.001,\, 0.01,\, 10,\, 10)$. ($Q^2$ in $\mathrm{GeV}^2$)
The distribution $F(b;\,0.001,\, 0.001,\, 10,\, 10)$ for two
subcollisions at midrapidity is included for comparison.}
\label{fig:non-central}
\end{figure}

\begin{figure}
\centering
\epsfig{file=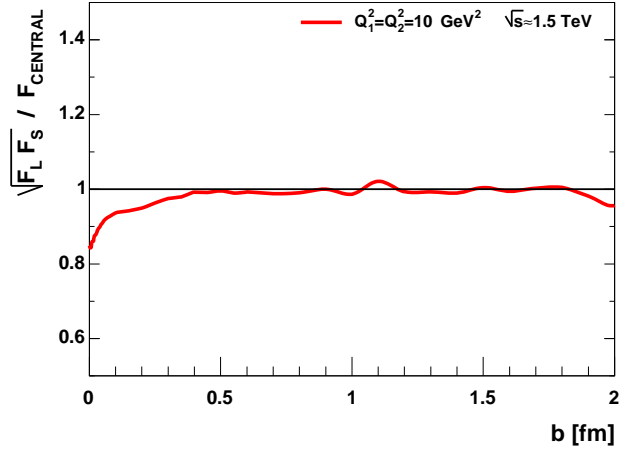,width=0.6\linewidth}
\caption{The ratio $\sqrt{F_L
    F_S}/F_{\mathrm{central}}$. $Q_1^2=Q_2^2=10$~GeV$^2$, and
  $\sqrt{s} \approx 1.5$~TeV.}
\label{fig:ratio-non-central}
\end{figure}

As mentioned in sec.~\ref{sec:formalism}, for subcollisions away from
midrapidity, the effective cross section is given by
\eqRef{eq:sigmaefflund}, which contains a product of two
$F$-distributions with different $x$-values. In
fig.~\ref{fig:non-central} we show results relevant for two
subcollisions with $Q^2=10$~GeV$^2$ at 1.5~TeV. One collision is at
rapidity zero, with $x_1=x_1'=0.001$, while the other subcollision has
$x_2=0.01$ and $x_2'=0.0001$, which corresponds to a rapidity for the
pair of jets (or a produced massive particle) given by
$y_{\mathrm{pair}}=\ln(x_2/x_2')/2\approx 2.3$. The corresponding
$F$-distributions are called $F_S$ (for small $x_2=0.0001$) and
$F_{L}$ (for large $x_2=0.01$). We see that in the tail $F_S$ and
$F_L$ lie on opposite sides of the distribution
$F_{\mathrm{central}}$, which is the relevant one for two
subcollisions at midrapidity (both with $x_i=x_i'=0.001$). As
$\sigma_{\mathrm{eff}}$ is given by $(\int d^2b F_S F_L)^{-1}$, this
implies that although the $F$-distributions vary with $x$,
$\sigma_{\mathrm{eff}}$ is approximately independent of
$y_{\mathrm{pair}}$. This result is further illustrated in
fig.~\ref{fig:ratio-non-central}, which shows the ratio $\sqrt{F_S
  F_L}/F_{\mathrm{central}}$. We see that this ratio is close to 1
except for small $b$-values, which have a low weight in the integral
over $d^2b$.  As a consequence also the result for
$\sigma_{\mathrm{eff}}$ varies very little with rapidity, in this case
from 35.3 to 37.1~mb when $y_{\mathrm{pair}}$ is changed from 0 to
2.3 (see table \ref{table:results}).  Actually, for small shifts
$\delta y_{\mathrm{pair}}$, the difference between the product $F_L
F_S$ and $F_{\mathrm{central}}^2$ must be only second order in $\delta
y_{\mathrm{pair}}$. This must then also be the case for the
corresponding values for $\sigma_{\mathrm{eff}}$.

\subsection{Comparison with experiment}

The result for $\sigma_{\mathrm{eff}}$ is larger than the effective
cross section observed for $\gamma+3$jet events at the Tevatron.  We
note, however, that these data depend on quark distributions, and not
only on the gluon distributions. It would be interesting to try to
estimate the difference, but this goes beyond the scope of this paper.
For $Q^2= 10^3\,\mathrm{GeV}^2$ at 1.5~TeV we find
$\sigma_{\mathrm{eff}}=23\,\mathrm{mb}$. Within the errors this is in
agreement with the 4-jet results from CDF, which however, have large
uncertainties. It would also be interesting to have, for comparison,
the values for $\sigma_{\mathrm{eff}}$ in the tuned versions of
\pythia, not only for the less successful single Gaussian
approximation used in section \ref{sec:pythiaimpact}.

We also note that there are two observed features, which are well
reproduced by our model:

\emph{i}. The D0 collaboration \cite{Abazov:2009gc} has observed a
reduction in $\sigma_{\mathrm{eff}}$ for growing $p_\perp$ of the jet
pair in $\gamma+3\,$jet-events. Although the errors are large, the
central values for $\sigma_{\mathrm{eff}}$ drop from 18.2 to 13.9~mb,
when $\langle p_T^{\mathrm{jet2}}\rangle$ increases from 17.6 to
27.3~GeV. This agrees qualitatively with the variation with $Q^2$
found in our model.

\emph{ii}. The CDF collaboration \cite{Abe:1997xk} observes in the same
reaction, that 
$\sigma_{\mathrm{eff}}$ is very insensitive to a change in the rapidity 
$y_{\mathrm{pair}}$ (or equivalently $x_F^{\mathrm{pair}}$) for one of the 
subcollisions. As discussed in sec.~\ref{sec:non-central} this is also consistent
with our model.  We note, however, that this result is mainly a consequence
of the fact that the product $F_L\,F_S$ is close to $F_{\mathrm{central}}^2$, and
does not neccesarily show that $F$ does not vary with $x$.

These qualitative agreements with experimental data may indicate that,
also if our model possibly underestimates the correlations, we may
believe that the qualitative features, and variations with $Q^2$ and
$\sqrt{s}$, are correct. The indicated strong dependence on $Q^2$
should then be very important for the interpretation of multiple hard
interactions at LHC. We here note a possible qualitative difference
between our model and the model with $x$-dependent matter densities in
ref.~\cite{Corke:2011yy}. Our result shows a larger variation with
$Q^2$ for fixed energy, than with $x$ (or energy) for fixed $Q^2$. At
fixed energy a variation with $Q^2$ is equivalent to a variation with
$x=Q/\sqrt{s}$. However, the two models may give different results
when data at different energies are compared. We would like to study
this question more in the future.

\subsection{Comment on the definition of $b$}

\begin{figure}
\centering
\epsfig{file=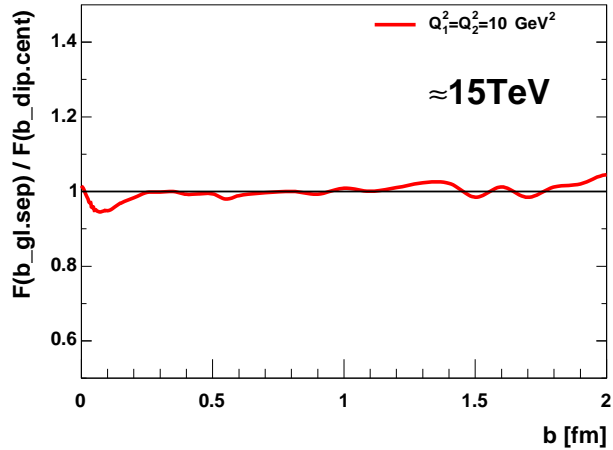,width=0.6\linewidth}
\caption{The ratio between the $F$-distributions obtained by two different
  definitions of $b$. $Q_1^2=Q_2^2=10$~GeV$^2$, and $\sqrt{s} \approx 15$~TeV.}
\label{fig:bdef}
\end{figure}

As mentioned in sec.~\ref{sec:application}, in the dipole model $F$ depends 
in principle on three vectors: the size of the two dipoles ($\mathbf{r}_1$,
$\mathbf{r}_2$) and the separation between them. For the results presented
above we have defined the separation as the distance between two gluons, and
averaged over $\mathbf{r}_1$ and $\mathbf{r}_2$. When calculating $\int d^2 b
\,F(b)$ this gives exactly the correct result. This is, however, not the case
for the integral $\int d^2 b\, F^2(b)$, which determines $\sigma_{\mathrm{eff}}$.
When the separation is of the same order of magnitude as the size of the 
dipoles, the averaging does not give the correct result. 
To estimate this error we have changed the definition of $\mathbf{b}$, to the
separation between the centers of the dipoles, keeping this fixed when
averaging over angles for $\mathbf{r}_1$ and $\mathbf{r}_2$. The ratio between 
the two different $F$-distributions is shown in fig.~\ref{fig:bdef}. 
In this example we have $Q_1^2=Q_2^2=10 \,\mathrm{GeV}^2$ and $\sqrt{s} \approx 15$~TeV.
We can here see a difference when $b$ is of the same
order of magnitude as the dipoles, around 0.1~fm. The contribution
from small $b$-values is, however, suppressed in the integral over $d^2b$,
and as mentioned above $\int d^2b \,F^2(b)$ obtains its major contribution for 
$b$ in the range 0.5-1~fm. The difference between the two estimates of  
$\sigma_{\mathrm{eff}}$ is only about 2.5\%. For larger $Q^2$ the difference is
even smaller, as the region where the two $b$-definitions give different 
results moves towards smaller $b$-values. This effect can therefore be safely
neglected.

\section{Conclusions and outlook}
\label{sec:conclusions}

The Lund Dipole Cascade model offers unique possibilities to study the
evolution of gluons inside hadrons at small $x$. The formalism is
based on BFKL evolution including essential higher-order corrections
and saturation effects. By following the evolution,
emission-by-emission, in rapidity and in transverse position, we can
investigate the correlations and fluctuations of the gluon
distribution in great detail.

We have here concentrated on the double parton distribution, which 
enters into the multiple parton scattering cross section in proton-proton
collisions. The double parton distribution in transverse coordinate space is
the analogy of the generalized parton density in momentum space.
The correlations can be described by a distribution in impact parameter space, 
$F(b;x_1, x_2, Q_1^2, Q_2^2)$. In many applications this distribution is 
assumed to be independent of the energy fractions and scales
of the partons, and the same for quarks and gluons. In the
\pythia event generator the width of the distribution does vary with energy, and
in a recent study it is assumed to vary with $x$. Earlier analyses have
demonstrated that DGLAP evolution implies non-trivial correlations depending
on $x$ and $Q^2$, but this analysis does not give information about the
$b$-dependence. In our analysis we find that the two-parton correlation
function $F(b)$ depends in a non-trivial way on all the kinematic variables 
$x_1,\,x_2,\,Q_1^2,\,Q_2^2$. 

For subcollisions at \emph{midrapidity}, $F$ is directly
connected to the ``effective cross section'' used in experimental analyses, via
the relation \mbox{$\int d^2b \,F^2=\sigma_{\mathrm{eff}}^{-1}$}. The result
that $F$ depends on $x_i$ and $Q_i^2$  implies, however, that for
subcollisions away from $y=0$, $\sigma_{\mathrm{eff}}$ is determined by an
integral of a product of two $F$-functions with different $x$-values, 
connected by the relation $x_i x_i'=Q^2/ s$. Therefore $\sigma_{\mathrm{eff}}$
is very insensitive to the rapidity of a hard subcollision; when one of the
two $F$-functions in the product is above, the other is below the
$F$-distribution relevant for midrapidity subcollisions. An experimental
result showing a weak dependence on rapidity is therefore not a proof that $F$
does not depend on the scaling parameters $x_i$.

In our formalism we can also study
event-by-event fluctuations in the density of partons, which are usually
neglected in other analyses. Neglecting the fluctuations $F$ satisfies the
normalization condition \mbox{$\int d^2b \,F=1$}, but when taking them into 
account we obtain a larger value for this integral, and therefore larger
two-parton correlations. In our analysis this effect
contributes 10-20\% to the value of $\sigma_{\mathrm{eff}}$. 

By studying the correlation function $F$, we can see explicitly how 
the emission of \mbox{high-$p_\perp$} gluons result in so-called
\emph{hot spots} developing for small $b$-values at large $Q^2$.
For fixed $Q^2$ the emission of larger dipoles, related to low-$p_\perp$ gluons,
implies that the distribution widens as we go down in $x$. (This effect is
related to the shrinking of the diffractive peak at higher energies.) As a 
result $\sigma_{\mathrm{eff}}$ is reduced for higher $Q^2$ at fixed energy, but
increases at higher energy for fixed $Q^2$. We see, however, that the result
appears to depend much stronger on $Q^2$ than on $x$ or collision energy.

It is not straight-forward to compare our results for the effective
cross section with data from the Tevatron and ISR, which correspond to lower
values for $\sigma_{\mathrm{eff}}$ and thus stronger correlations. The 
ISR data are obtained for very
high $x$-values, and the more exact Tevatron results are for $\gamma +3\,$jet
events, which necessarily involves quarks. Our calculations include
only gluons, and should therefore only be applied at small $x$. It is also
difficult to 
estimate a possible difference between quark and gluon distributions. 
Our results are, however, compatible within errors with the four-jet 
measurement at CDF, where gluons should be dominating, and they 
show qualitative agreements with the
observed rapidity independence at CDF and the hints of strong
$Q^2$-dependence at D0.
We also note that our results appear to be in qualitative agreement with
results from \pythia tuned to multiple interactions and underlying events.

Even if it is possible that we overestimate the result for 
$\sigma_{\mathrm{eff}}$ to some degree, we are more confident about the
variation with $x$ and $Q^2$, which should have relevance for extrapolations
to the LHC. Our model here predicts a
very strong decrease of the effective cross section with increasing
$Q^2$ (especially when both $Q^2_1$ and $Q^2_2$ increases), while the
dependence on the total energy ($\sim 1/x$) for fixed $Q^2$ is rather
weak. Also the dependence on the rapidity of the two subcollisions is 
quite weak, which follows as $\sigma_{\mathrm{eff}}$ here is given by a product
of two $F$-distribution, one which is smaller and one which is larger than the
$F$-distribution relevant for midrapidity subcollisions.

When comparing to the recent option in \pythia, with an $x$-dependent
impact parameter profile \cite{Corke:2011yy}, our result may be quite similar 
if we study the
dependence on $Q^2$ at a fixed cms energy $\sqrt s$. In our model the
correlations increase as a direct consequence of higher $Q^2$, and in the 
\pythia model they increase
because $x$ is higher for large $Q^2$ at fixed energy. The difference between
the two models might therefore show up in the extrapolation of Tevatron data
to LHC energies.

It should be noted, however, that the extraction of the effective
cross sections from data is very hard, and it will still be difficult
to compare future LHC results to the numbers we have presented
here. For this reason it is important to compare to event generator
predictions for the experimental observables. Fortunately we have now
developed the \dipsy MC to also generate exclusive final states
\cite{Flensburg:2011kk}, which means that in the future we can in
detail simulate exclusive final state observables related to double parton
scattering.

\section*{Acknowledgments}

Work supported in part by the EU Marie Curie RTN MCnet
(MRTN-CT-2006-035606), and the Swedish research council (contracts
621-2008-4252 and 621-2009-4076).

L.L.\ gratefully acknowledges the hospitality of the CERN theory unit.

A.S.\ gratefully acknowledges the support of
   the Hungarian OTKA grants T49466 and NK 73143.

\bibliographystyle{utcaps}  
\bibliography{doublescat} 

\end{document}